\documentclass{aa} 
\usepackage{graphics} 
\newcommand{\chem}[2]{$\rm{}^{#1}\kern-0.8pt#2$}
\newcommand{\chim}[2]{\rm{}^{#1}\kern-0.8pt#2} 
\newcommand{\reac}[6]{$\rm\,{}^{#1}\kern-0.8pt{#2}\,({#3}\,,{#4})\, {}
^{#5}\kern-0.8pt{#6}\,$} 

\begin{document}

\thesaurus{ 1(02.14.1;08.01.1;08.19.4;07.19.2)}

\title{The synthesis of the light Mo and Ru isotopes: how now, no need for an exotic
solution ?}

\author{} \author{V. Costa\inst{1,2,3} \and M. Rayet\inst{3} 
\and R. A. Zappal\`a\inst{1}
\and M. Arnould\inst{3}} \institute{Dipartimento di Fisica e Astronomia dell'Universit\`a
degli studi di Catania, Italy \and INFN-LNS, Catania, Italy \and Institut
d'Astronomie et d'Astrophysique, Universit\'e Libre de Bruxelles, Belgium}
\titlerunning{The synthesis of the light Mo and Ru isotopes}
\date{Received date; accepted date} \maketitle

\begin{abstract}
The most detailed calculations of the p-process call for its development in the O/Ne
layers of Type II supernovae. In spite of their overall success in reproducing the
solar system content of p-nuclides, they suggest a significant underproduction of the
light Mo and Ru isotopes. On grounds of a model for the explosion of a 25
$M_{\odot}$ star with solar metallicity, we demonstrate that this failure might just be
related to the uncertainties left in the rate of the \reac{22}{Ne}{\alpha}{n}{25}{Mg}
neutron producing reaction. The latter indeed have a direct impact on the distribution of
the s-process seeds for the p-process. 
 
\end{abstract}

\section{Introduction}
\label{introduction}

The most successful p-process models available to-date call for the synthesis of the
stable neutron-deficient nuclides heavier than Fe in the O/Ne layers of Type II
supernovae (SNII) (Rayet et al. 1995, hereafter RAHPN). In spite of their many virtues
in reproducing the solar-system p-nuclide abundance distribution, they however suffer
from some shortcomings. One of them concerns their persistent underproduction of the
light Mo (\chem{92}{Mo}, \chem{94}{Mo}) and Ru (\chem{96}{Ru}, \chem{98}{Ru}) p-isotopes.
Some have tried to remedy this situation with exotic solutions, calling in
particular for accreting neutron stars or black holes (e.g. Schatz et al. 1998). The 
level of the contribution of such sites to the solar system content of the nuclides of 
concern here is quite impossible to assess in any reliable way. 
In contrast, it has been emphasized many times over the last decade
that the problem might just be due to some misrepresentation of the production in
the He-burning core of massive stars of the s-nuclide seeds for the p-process (e.g.
Arnould et al. 1998).

The aim of this Letter is to scrutinize the latter, `non-exotic', solution in a
quantitative way by duly taking into account the uncertainties that still affect the
rate of the
\reac{22}{Ne}{\alpha}{n}{25}{Mg} reaction, as they appear in the
NACRE compilation of reaction rates (Angulo et al. 1999). Clearly, these uncertainties
in the key neutron producer in conditions obtained during central He burning in
massive stars have a direct impact on the predicted abundances of the s-nuclide seeds
for the p-process, as already analyzed quantitatively by Meynet \& Arnould (1993).
Another potential embarassment of the p-process predictions identified by RAHPN is a
SNII overproduction of oxygen relative to the p-nuclides. We show that this problem
might be cured along with the one of the underproduction of the
light Mo and Ru isotopes if the \reac{22}{Ne}{\alpha}{n}{25}{Mg} rate
is modified adequately within a range permitted by the NACRE compilation.  
For the sake of illustration, we just discuss here the case of a 25 M$_{\odot}$ solar
metallicity ($Z = Z_{\odot}$) star. A more complete
study dealing in particular with a set of stars with different masses and
metallicities, and analyzing the impact of the uncertainties in the rates of a variety
of reactions, is currently under way.

The adopted input physics is briefly described in Sect. 2, and some results are
presented in Sect. 3. Conclusions are drawn in Sect.~4.

\section{Input physics}
 
The calculations reported here are based on a model for a $Z_{\odot}$ 8
M$_{\odot}$ helium star already considered by RAHPN. It corresponds to a
main sequence mass of about 25 M$_{\odot}$ and is  evolved from the beginning of core
helium burning to the supernova explosion. Details about this model can be found in 
Hashimoto (1995), and are summarized in RAHPN. As in
RAHPN, 20 O/Ne-rich layers with explosion temperatures peaking in the 
(1.8-3.3)$\times 10^9$ K range are selected as the P-Process Layers (PPLs).
Their total mass is approximately 0.58~M$_{\odot}$. The deepest PPL is
located at a mass of about 1.94~M$_{\odot}$, which is far enough from the mass cut for all
the nuclides produced in this region to be ejected during the explosion. 
 
The p-process reaction network and its
numerical solver are described by RAHPN. A series of their selected
nuclear reaction rates are updated, however. In particular, the NACRE `adopted'
rates are used for charged particle captures by nuclei up to \chem{28}{Si}.  For
heavier targets, the rates predicted by the Hauser-Feshbach  code
MOST (Gor\-iely 1997) are used, except for the experimentally-based neutron capture rates
provided by Beer et al. (1992).\footnote{The NACRE and MOST rates are available in the
Brussels Nuclear Astrophysics Library (http://www-astro.ulb.ac.be)}
 
As already pointed out in Sect. 1, we turn our special attention to
the impact of the uncertainties remaining in the rate of
\reac{22}{Ne}{\alpha}{n}{25}{Mg}. For temperatures of about $2-3 \times 10^8$ K at
which the s-process typically develops during core He burning in massive stars
(e.g. Rayet \& Hashimoto 2000), the NACRE upper limit on this rate is
50-500 times larger than the `adopted' value (see Angulo et al. 1999 for
details). In order to quantify the consequences of this situation for the
predicted abundance distribution of the s-nuclide seeds for the p-process, and ultimately
for the p-nuclide yields themselves, we perform nucleosynthesis calculations for five
different rates ranging from the NACRE adopted value to its upper limit. These rates,
labelled R$_i$ ($i$=1 to 5) in the following, are defined and displayed in Fig.~1. 
\footnote{The calculations reported here were completed when we have been informed of new
low-energy measurements of the \reac{22}{Ne}{\alpha}{n}{25}{Mg} cross section  (J.W. Hammer
\& M. Jaeger, private communication). Rate estimates based on these new data are not
available yet. It seems, however, that the revised upper limit might range somewhere
between R$_4$ and R$_5$} 
They are used in the $Z_{\odot}$ 25 M$_\odot$ star referred to above to calculate the
abundances of the s-process nuclides at the end of core He burning. The results are shown in
Fig.~2 for the s-only nuclides.  
Use of R$_1$ leads to the classical `weak' s-process component pattern (e.g. Rayet \&
Hashimoto 2000), exhibiting a decrease of the overproduction (with respect to solar) of
the s-nuclides by a factor ranging from $\approx$100  to about unity when the  mass
number $A$ increases from about 70 to 100. In the heavier mass range, the s-process
`main component' supposed to originate from low- or intermediate-mass stars takes
over. This `canonical' picture changes gradually with an increase of the
\reac{22}{Ne}{\alpha}{n}{25}{Mg} rate, more \chem{22}{Ne} having time to burn,
releasing more neutrons, before He exhaustion in the core. The
direct result of this is a steady increase of the overproduction of heavier and heavier
s-nuclides. For example, with the extreme R$_5$ rate, the overproduction factor increases
from $10^3$ to $10^4$ for $A$ varying from about 70 to 90, before decreasing to
a value around unity for $A \approx 150$ only.

\begin{figure}
\resizebox{\hsize}{!}{\rotatebox{270}{\includegraphics{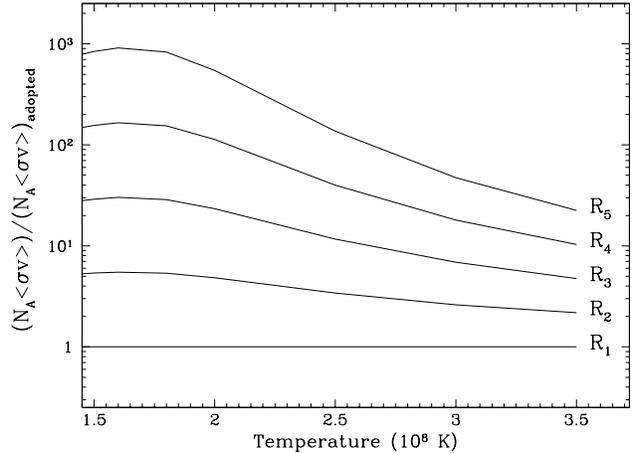}}}
\caption{
The five \reac{22}{Ne}{\alpha}{n}{25}{Mg} rates R$_i$ used in our calculations for 
temperatures of relevance for core He burning in
massive stars. All rates are normalized to R$_1$, which is the NACRE `adopted'
rate. R$_5$ corresponds to the NACRE upper values. R$_3$, R$_2$ and R$_4$ are 
the geometrical means between R$_1$ and R$_5$, R$_1$ and R$_3$, and R$_3$ and R$_5$, 
respectively (see also footnote 2)}

\end{figure}

 \begin{figure}
 \resizebox{\hsize}{!}{\rotatebox{270}{\includegraphics{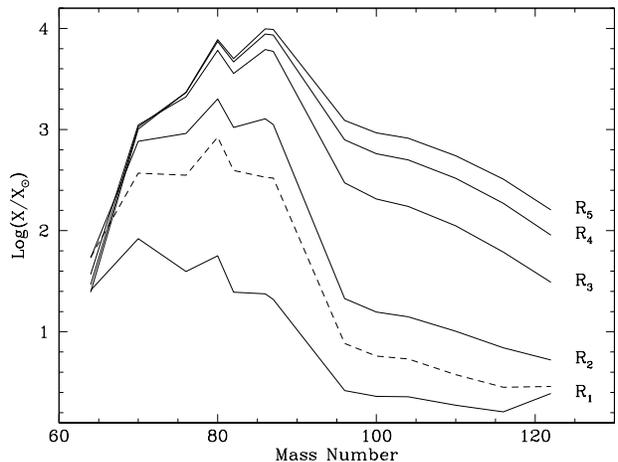}}}
 \caption{
 Distribution of the abundances, normalized to solar values, of the s-only
 nuclides at the end of core He burning in the considered $Z_{\odot}$ 25
 M$_\odot$ model star, for the  \reac{22}{Ne}{\alpha}{n}{25}{Mg} 
 rates R$_i$ (i = 1 to 5) defined in Fig.~1, all the other ingredients of 
 the model being kept unchanged (see Rayet \& Hashimoto 2000).  The dashed 
 line is the distribution adopted by RAHPN for their p-process calculations}
  
 \end{figure}

At first sight, it might be felt that the s-process abundance distributions obtained
with large enough R$_i$ values exhibit some unwanted or embarrassing features. One of
these concerns the underproduction of the $A \approx 70-76$ s-nuclides relative
to the $A \approx 80-90$ ones. Another one relates to the fact that a more or less
substantial production of heavy s-nuclides (like in the Ba region) would screw up the
pattern of the s-process main component ascribed to lower-mass stars. In our
opinion, none of these predictions can really act as a deterrent to
\reac{22}{Ne}{\alpha}{n}{25}{Mg} rates substantially in excess of R$_1$. On the one
hand, the absence of ab initio self-consistent calculations of the s-process in low-
and intermediate mass stars does not allow at this time to predict the exact shape of
the main s-process component which is classically assigned to these stars. As a
consequence, a contribution to the main component by massive stars cannot be excluded,
even if it may disturb some traditional views on the subject.  On the other hand,
the reduction of the light s-process nuclide production by massive stars could well be
compensated by their increased synthesis by some low- or intermediate-mass stars when
rates larger than R$_1$ are considered (Goriely \& Mowlavi 2000). The classical
\chem{80}{Kr} overproduction problem found in the massive star s-process
(e.g. Rayet \& Hashimoto 2000) could also be eased with increased 
\reac{22}{Ne}{\alpha}{n}{25}{Mg} rates, as demonstrated by Fig.~2. For these same
rates, note that \chem{80}{Kr} is not overproduced either in some of the calculations of
Goriely \& Mowlavi (2000) which predict high yields of the other light s-nuclides.  

Figure~2 also suggests that a discrepancy, if any, between the observed Ba overabundance in
the SN1987A ejecta and the model
predictions  could be cured in a natural way by increasing the adopted
\reac{22}{Ne}{\alpha}{n}{25}{Mg} rate. The  
[Ba/Fe]$_{SN}$/[Ba/Fe]$_{LMC}$ ratio is observationally still quite uncertain, values
between about 5 and 20 having been reported (e.g. Mazzali \& Chugai 1995). Prantzos et
al. (1988) have calculated lower values of 2.6 to 4.7 with the
\reac{22}{Ne}{\alpha}{n}{25}{Mg} rate of Fowler et al. (1975). This rate is on average
comparable to R$_1$ in the temperature range of relevance to the s-process. We
have not conducted any new s-process calculation in a specific SN1987A progenitor model.
Instead, some rough estimates based on the procedure of Prantzos et al. (1988) in which
their adopted s-process Ba mass fraction is replaced by the one calculated for the model
star adopted here have been made. Assuming that the LMC metallicity is one third of the
solar one, we predict [Ba/Fe]$_{SN}$/[Ba/Fe]$_{LMC}$ ratios from 3 to 14 for rates 
increasing from R$_1$ to R$5$. Theory could thus account for quite substantial 
SN1987A Ba productions with high enough \reac{22}{Ne}{\alpha}{n}{25}{Mg} 
rates (compatible with the NACRE data). 

As discussed by RAHPN, it is a fair approximation to adopt the s-process 
abundance distributions of Fig.~2 as seeds for the p-process. For the 
$A \le 40$ species,  the initial abundances in the PPLs are taken from the 
detailed stellar models. Although these models have been obtained with
rates that may differ from the NACRE ones adopted here, this 
inconsistency is certainly not responsible for any intolerable distorsion 
in the predicted s-process seeds or p-process yields.

\section{Results and discussion}
\label{results}

The various seed abundances of Fig.~2 are used to compute the production of
the p-nuclides in the PPLs of the  $Z_{\odot}$
25 M$_\odot$ star considered here. As in RAHPN, the abundance of a p-nuclide $i$ is 
characterized by its mean overproduction factor 
$\left< F_i \right> = \left< X_i \right> /X_{i,\odot}$, where $X_{i,\odot}$ is its
solar mass fraction (Anders \& Grevesse 1989), and
\begin{equation}
\left<X_i\right> = 
\frac{1}{M_p} \sum_{n \geq 1} (X_{i,n} + X_{i,n-1}) (M_n - M_{n-1})/2,
\end{equation}
where $X_{i,n}$ is the mass fraction of isotope $i$ at the mass coordinate $M_n$,
 $M_p = \sum_{n\geq 1} (M_n - M_{n-1})$ is the total mass of the PPLs, the sum running
over all the PPLs ($M_0$ corresponds to the bottom layer).  An overproduction factor
averaged over all 35 p-nuclei is calculated as $F_0 = \sum_i \left< F_i \right>/35$, and is
a measure of the global p-nuclide enrichment in the PPLs. So, if the computed p-nuclei
abundance distribution were exactly solar, the normalized mean overproduction factor
$\left< F_i \right>/F_0$ would be equal to unity for all $i$.
 
\begin{figure}
\resizebox{\hsize}{!}{\includegraphics{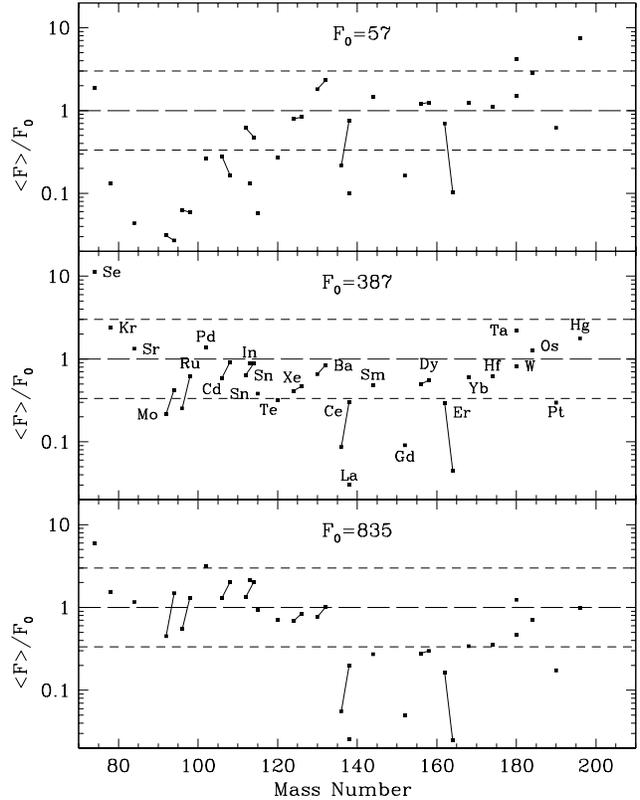}}
\caption{
Values of $\left< F_i \right>/F_0$  and of $F_0$ derived from the seed
abundances calculated with the \reac{22}{Ne}{\alpha}{n}{25}{Mg} rate R$_1$ (upper panel),
R$_3$ (middle panel) and R$_5$ (lower panel). Lines connect different isotopes of the same
element.  The dotted horizontal lines delineate the $0.3 \leq \left< F_i
\right>/F_0
\leq 3$ range}
 
\end{figure}
  
Figure 3 shows the
normalised p-nuclide overproduction factors derived from the seed abundance
distributions calculated with the \reac{22}{Ne}{\alpha}{n}{25}{Mg} rates R$_1$, R$_3$ and
R$_5$. Changes in the shape of the p-nuclide abundance distribution are clearly noticeable,
at least for $A \la 120$. The use of
R$_1$ leads to a more or less substantial underproduction of not only \chem{92}{Mo},
\chem{94}{Mo},
\chem{96}{Ru} and \chem{98}{Ru}, a `classical' result in p-process studies (see RAHPN), but
also of
\chem{78}{Kr} and \chem{84}{Sr}, which was not predicted in previous calculations. This new
feature directly relates to the larger abundances around $A \approx 80$ used by RAHPN
(dashed curve in Fig.~2), in contrast to the much flatter seed distribution 
obtained with R$_1$. This Kr-Sr-Mo-Ru trough is gradually reduced, and in fact 
essentially disappears, for
\reac{22}{Ne}{\alpha}{n}{25}{Mg} rates of the order or in excess of R$_3$. This situation
is most clearly illustrated by Fig.~4. In these very same conditions,   
$\left< F_i \right>/F_0$ for \chem{113}{In} and \chem{115}{Sn} comes much closer to
unity as well. It has to be noticed that this
situation does not result from a stronger production of these two nuclides by the
p-process, but instead from their increased initial abundances associated with a more
efficient  s-process when going from R$_1$ to R$_5$. In contrast, the 
$\left< F_i \right>/F_0$ pattern does not depend on the
adopted \reac{22}{Ne}{\alpha}{n}{25}{Mg} rate for $A \ga 140$. This is expected from a mere
inspection of the s-nuclide seed distributions displayed in Fig.~2. In particular, 
\chem{152}{Gd} and
\chem{164}{Er} remain underproduced. This cannot be considered as an embarassment as these
two nuclides can emerge from the s-process in low- or intermediate-mass stars.
 
\begin{figure}
\resizebox{\hsize}{!}{\rotatebox{270}{\includegraphics{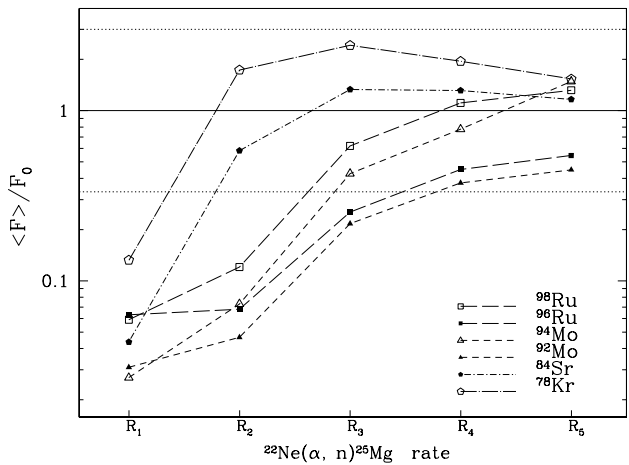}}}
\caption{
Values of the normalized overproduction factors of the Kr, Sr, Mo and Ru 
p-isotopes as a function of the \reac{22}{Ne}{\alpha}{n}{25}{Mg} rate}
 
\end{figure}

In addition, the overall efficiency of the p-nuclide production substantially
increases with increasing \chem{22}{Ne} burning rates. More specifically, $F_0$
is multiplied by a factor of about 15 when going from R$_1$ to R$_5$. This could largely
ease, and even solve, the problem of the relative underproduction of the p-nuclides
with respect to oxygen identified by RAHPN. For their considered 25 M$_{\odot}$ model
star calculated with the \reac{12}{C}{\alpha}{\gamma}{16}{O} rate from Caughlan et al.
(1985), they obtain $F_0=130$ and report a value of 4.4 for the ratio of the oxygen 
to p-process yields. This value would come close to unity for  
\reac{22}{Ne}{\alpha}{n}{25}{Mg} rates in the vicinity of R$_3$-R$_4$, 
as the p-nuclides would be about 3 to 6 times more produced than in RAHPN.  

\section{Conclusions}

This Letter makes plausible that the long-standing puzzle of the underproduction 
with respect to solar of the p-isotopes of Mo and Ru in SNII explosions could be
quite naturally solved by just assuming an increase of the
\reac{22}{Ne}{\alpha}{n}{25}{Mg} rate over its `nominal' value. More specifically,
this could be achieved by multiplying the NACRE `adopted' rate by factors of about
10 to 50 in the temperature range at which the s-process
typically develops during core He burning in massive stars. These factors are well
within the uncertainties reported by NACRE. As an important bonus, this increased rate
would also largely avoid (i) the underproduction of
\chem{78}{Kr} and of \chem{84}{Sr} which we predict here for the first time to be
concomitant to the light Mo and Ru one, (ii) the too low production of \chem{113}{In}
and \chem{115}{Sn}, and (iii) the overall underproduction of the p-nuclides with respect to
oxygen noted by RAHPN. In direct relation with an increased
\reac{22}{Ne}{\alpha}{n}{25}{Mg} rate, more s-process Ba could also be ejected by SNII
events. Our predictions confortably overlap the range of Ba overabundances reported
for SN1987.

This array of pleasing features has of course not to be viewed as a proof of the
validity of the assumption that the true \reac{22}{Ne}{\alpha}{n}{25}{Mg} rate is
higher than usually thought. It may just be a hint that there might be ways around
exotic solutions. This conclusion applies at least if one relies on the simplistic (and the
only ones to be available for our purpose) supernova models used here and in previous
p-process calculations (see RAHPN et references therein), as well as in a myriad of other
explosive nucleosynthesis calculations.

\begin{acknowledgements}
R.Z. and V.C. thank the E.U. for financial support of the PhD in Physics at the
University of Catania through the FSE.
M.R. is Research Associate of the FNRS (Belgium).

\end{acknowledgements}

\end{document}